\documentclass[superscriptaddress,aps,pra,twocolumn,showkeys]{revtex4-2}
\usepackage{amsmath,mathtools,siunitx}
\usepackage{amssymb}
\usepackage{graphicx}
\usepackage{color}
\usepackage{bm}
\usepackage[colorlinks,linkcolor=blue,anchorcolor=blue,citecolor=blue,urlcolor=blue]{hyperref}
\usepackage{booktabs}
\usepackage{braket} 
\usepackage{bm}

\begin{document}
\title{Entanglement transfer between giant atoms in waveguide-QED systems}

\author{Jie Liu}
\affiliation{Lanzhou Center for Theoretical Physics, Key Laboratory of Theoretical Physics of Gansu Province, and Key Laboratory of Quantum Theory and Applications of MoE, Lanzhou University, Lanzhou, Gansu 730000, China}

\author{Zhi-Qiang Liu}
\affiliation{Lanzhou Center for Theoretical Physics, Key Laboratory of Theoretical Physics of Gansu Province, and Key Laboratory of Quantum Theory and Applications of MoE, Lanzhou University, Lanzhou, Gansu 730000, China}

\author{Yu Sang}
\affiliation{Lanzhou Center for Theoretical Physics, Key Laboratory of Theoretical Physics of Gansu Province, and Key Laboratory of Quantum Theory and Applications of MoE, Lanzhou University, Lanzhou, Gansu 730000, China}

\author{Lei Tan}
\email{tanlei@lzu.edu.cn}
\affiliation{Lanzhou Center for Theoretical Physics, Key Laboratory of Theoretical Physics of Gansu Province, and Key Laboratory of Quantum Theory and Applications of MoE, Lanzhou University, Lanzhou, Gansu 730000, China}

\begin{abstract}
We investigate the entanglement transfer between giant atoms in waveguide-QED systems. The system consists of two pairs of two-level giant atoms, $ab$ and $cd$, each independently coupled to its respective one-dimensional waveguide. Initially, entangled states are stored in atom pair $ac$. There we consider three giant atom coupling configurations: separated, braided, and nested. For comparison, the entanglement transfer for small atoms configuration is also studied here. We focus on the entanglement transfer from atom pair $ac$ to atoms pair $bd$ and atom pair $ab$ in these four coupling configurations. It is shown that the transfer of entanglement in each coupling configuration strongly depends on phase shift. In particular, the braided configuration demonstrates superior performance in entanglement transfer. For the entanglement transfer from atom pair $ac$ to atom pair $bd$, complete entanglement transfer is presented in braided configuration, a behavior not found in small atom or other giant atom configurations. For the entanglement transfer from atom pair $ac$ to atom pair $ab$, although the maximum entanglement transferred to atom pair $ab$ in the braided configuration is half of that one to atom pair $bd$, it is still higher than that in the small atom, separated, and nested configurations. This study lays the foundation for entanglement transfer between giant atoms in waveguide-QED platforms.

\end{abstract}

\maketitle
\section{INTRODUCTION}

\begin{figure*}[!htbp]
	\includegraphics[width=0.95\textwidth]{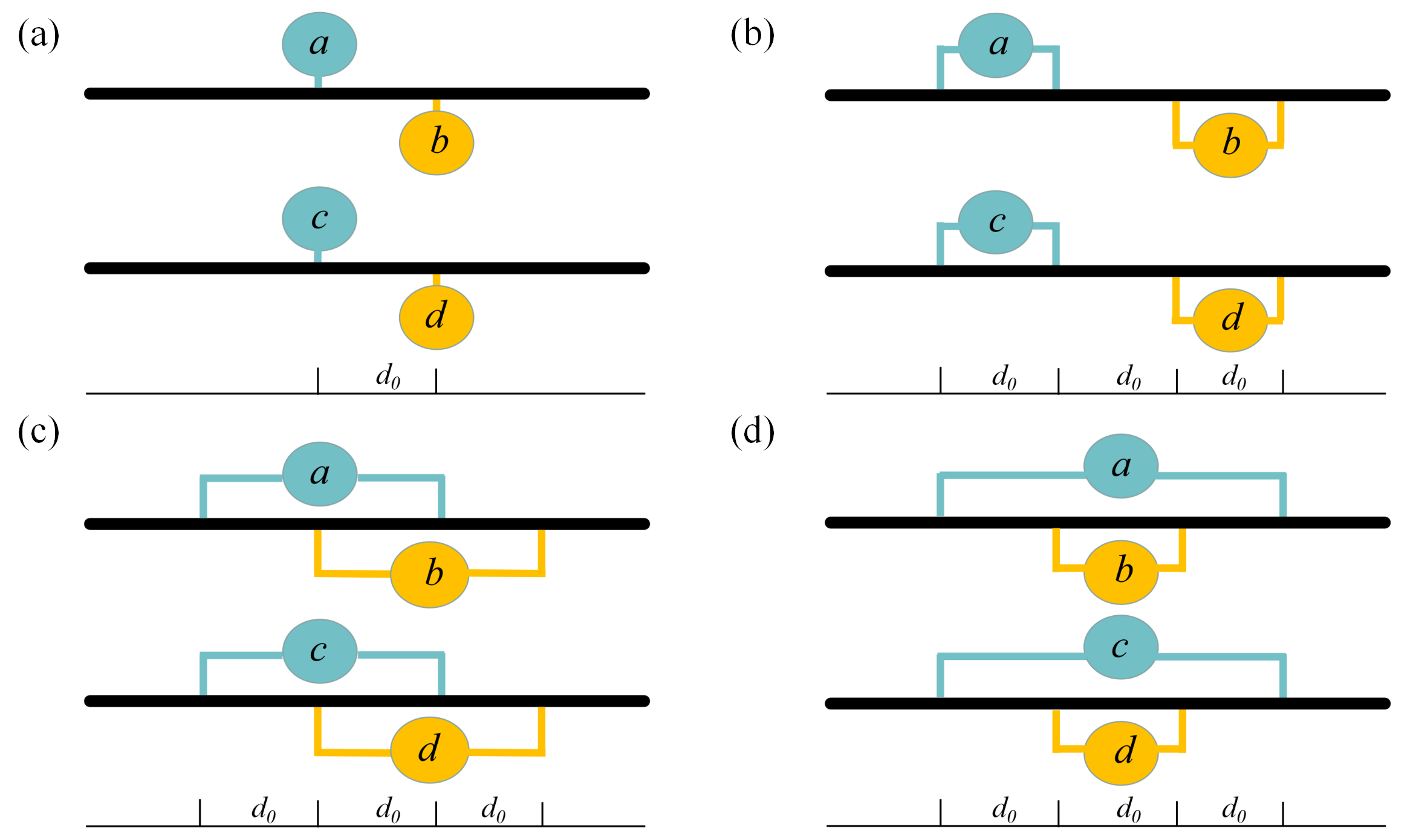}\hfill
	\caption{(Color online)  Schematic illustration of atom pair $ab$ and atom pair $cd$ coupled to their respective waveguides: $(a)$ small atom configuration, $(b)$ separated configuration, $(c)$ braided configuration, $(b)$ nested configuration. Configurations in $(b)$, $(c)$, and $(d)$ represent giant atom setups, where each giant atom couples to the waveguide at two connection points. In all panels, the atom pair $ac$ is initially chosen as the maximally entangled state, while the atom pair $bd$ is located in the ground state. For simplicity, the distance between two adjacent connection points is assumed to be the same, denoted by $d_0$, and the phase shift between them is $\phi = k_0 d_0 $ (where $ k_0 $ is the wave vector).
		Another assumption we adopt is that the frequencies of all four atoms are equal, i.e., $\omega_a = \omega_b = \omega_c = \omega_d$.}
	\label{fig1}
\end{figure*}
Waveguide quantum electrodynamics (QED) has attracted considerable attention due to its strong coupling with emitters \cite{Roy89, Gu2017} and its potential as a platform for quantum information processing \cite{Zheng111, Paulisch2016}. Many advancements have been made in waveguide-QED systems, including the single-photon \cite{Zhou101, Liao92, Shen95, Liao80, Liao81} and few-photon transport \cite{Trivedi98, Stolyarov99, Joanesarson101}, the creation of subradiant and superradiant states \cite{DeVoe76, Zanner2022, vanLoo2013}, as well as atomic entanglement generation \cite{Gonzalez106, Cano84, Tudela110, Ballestero89, Zheng110, Facchi2016}. In these studies of waveguide-QED systems, the atom-waveguide coupling is assumed to be point coupling because the size of the atom is much smaller than the wavelength of the propagating light associated with the atomic transition frequency in the waveguide. 

Recently, giant atoms, which are artificial atoms, have been introduced into waveguide-QED systems. Unlike the single point coupling of small atoms with waveguide, giant atoms can be coupled to a waveguide at multiple points \cite{frisk2021mathematics}. Many noteworthy physical phenomena have been reported in giant atom waveguide-QED systems to date, such as frequency-dependent relaxation rates and Lamb shift \cite{Kockum46}, waveguide-mediated decoherence-free interactions \cite{Kockum47, Kannan48, Cilluffo49, Carollo50, Soro51}, chiral and nonreciprocal scattering \cite{Zhou107, Zheng1092024, Chen2022}. In addition, as a key resource for quantum networks, the quantum entanglement generation in giant atom waveguide-QED systems has recently begun to be studied \cite{Yin55, Yin56, 130, luo57, liu2024}. These reports indicate that the generation of giant atom entanglement is governed by the phase shifts accumulated between the two connection points. Compared to the small atom setup, the giant atom in waveguide-QED systems exhibits greater maximum entanglement.

Besides the entanglement generation, it is also important to investigate how quantum entanglement can be transferred between subsystems, as it plays a vital role in quantum teleportation \cite{Bennett70} and quantum key distribution \cite{Grosshans2003}. Much work has been reported on entanglement transfer dynamics in small atomic systems \cite{Serafini73, Cavalcanti74, Sainz76, Romero81, Muhammed82, Mok2, Bougouffa2012, Yonac2007, xu2009}. However, the entanglement transfer between giant atoms remains an unexplored topic.

To address this knowledge gap, we have thoroughly analyzed the entanglement transfer between giant atoms in waveguide-QED systems. We consider, here, each giant atom to be coupled to the waveguide at two distinct connection points. Three coupling configurations of giant atoms are determined according to the arrangement of connection points : separated, braided, and nested \cite{Kockum47}. Each coupling configuration consists of four giant atoms, identified as $a$, $b$, $c$, and $d$. Here, atom pair $ac$ is chosen as a maximally entangled state at the initial moment and atom pair $bd$ is in the ground state. Although entanglement transfer involving two pairs of small atoms has been studied in Ref. \cite{Cavalcanti74}, where the complete entanglement transfer can be achieved, the effect of atomic dissipation caused by the environment on entanglement transfer has not been taken into account. In practical physical systems, entanglement is fragile and can be easily disrupted by environmental factors. Therefore, in this work, we also study the entanglement transfer of the small atom configuration under dissipative environment.

We first consider the case where the entanglement stored in atom pair $ac$ is transferred to atom pair $bd$. It is found that the entanglement transfer in each configuration (including small atom configuration) is governed by the accumulated phase shift between two adjacent connection points. Notably, in the braided configuration, the entanglement in atom pair $ac$ can be completely transferred to atom pair $bd$. We note that the complete transfer of entanglement between two pairs of giant atoms in the braided configuration arises from the decoherence-free property of the giant atoms, which is different from the complete entanglement transfer achieved by neglecting atomic dissipation in Ref. \cite{Cavalcanti74}. For other configurations (small atom configuration, separated configuration, and nested configuration), the maximum entanglement values transferred to the atom pair $bd$ are lower than 0.5. 
In particular, for small atom configuration, compared to the results of Ref. \cite{Cavalcanti74}, we find that entanglement is partially lost to the environment in the case where atomic dissipation is considered.

We then discussed the transfer of entanglement from atom pair $ac$ to atom pair $ab$. In the braided configuration, however, the maximum entanglement value transferred to $ab$ is only 0.5, yet it remains higher than that in the small atom, separated, and nested configurations.

The structure of the remaining sections of this paper is as follows. In Sec. \ref{II}, we present a description of the physical system studied and the related theoretical background. In Sec. \ref{III} and Sec. \ref{IV}, we analyze the entanglement transfer process for the four atomic coupling configurations introduced in Sec. \ref{II}. The discussion of our results and the experimental feasibility of our model can be found in Sec. \ref{V}.
Finally, in Sec. \ref{VI}, we summarize the key research findings of this study.
\section{Physical models and theoretical descriptions}\label{II}
\subsection{Physical System and Quantum Master Equation}
In this work, we consider a four-qubit system comprising two pairs of two-level atoms, with the atoms labeled as $a$, $b$, $c$, and $d$. As shown in Fig. \ref{fig1}, the atom pair $ab$ is coupled to one waveguide, while the atom pair $cd$ is coupled to another waveguide. The atoms in Fig. \ref{fig1}$(a)$ are considered small, each coupled to the waveguide at a single connection point. Figs. \ref{fig1}$(b)$- \ref{fig1}$(d)$ illustrate giant atom setups, in which each giant atom is coupled to the waveguide at two distinct points. There are three coupling configurations based on the position of the connection points \cite{Kockum47}: separated configuration [Fig. \ref{fig1}$(b)$], braided configuration [Fig. \ref{fig1}$(c)$], and nested configuration [Fig. \ref{fig1}$(d)$].

Initially, the atom pair $ac$ is prepared in the maximally entangled state $({\left| e_a, g_c \right\rangle + \left| g_a, e_c \right\rangle})/{\sqrt{2}}$, while the atom pair $bd$ is in the ground state $\left| g_b, g_d \right\rangle$. The initial state vector for the four-qubit system is expressed as:
\begin{equation}
	\label{Istate}
	|\psi(0)\rangle = \frac{1}{\sqrt{2}} (| e_a, g_c \rangle + | g_a, e_c \rangle) \otimes | g_b, g_d \rangle.
\end{equation}

Tracing out the field modes in the vacuum waveguide, the quantum master equation for these four coupling configurations shown in Fig. \ref{fig1} can be described as \cite{Kockum47} (\(\hbar = 1\))
	\begin{equation}
		\begin{aligned}
			\dot{\rho} = & -i \left[ H_{ab}, \rho \right] +\Gamma_{a} \mathcal{D}[\sigma_{-}^{a}] \rho +\Gamma_{b} \mathcal{D}[\sigma_{-}^{b}] \rho   \\
			 &+ \Gamma_{ab} \left[ (\sigma_{-}^{a}\rho\sigma_{+}^{b} - \frac{1}{2} \left\{ \sigma_{+}^{b}\sigma_{-}^{a}, \rho \right\})+ \text{H.c.}\right] \\
			 & -i \left[ H_{cd}, \rho \right] +\Gamma_{c} \mathcal{D}[\sigma_{-}^{c}] \rho +\Gamma_{d} \mathcal{D}[\sigma_{-}^{d}] \rho   \\
			 &+ \Gamma_{cd} \left[ (\sigma_{-}^{c}\rho\sigma_{+}^{d} - \frac{1}{2} \left\{ \sigma_{+}^{d}\sigma_{-}^{c}, \rho \right\})+ \text{H.c.}\right],
		\end{aligned}
		\label{eq:me1}
	\end{equation}

where 
	\begin{equation}
		H_{ab}= \delta \omega_{a} \sigma_{+}^{a}\sigma_{-}^{a} + \delta \omega_{b} \sigma_{+}^{b}\sigma_{-}^{b} + g_{{ab}} \left(\sigma_{+}^{b}\sigma_{-}^{a} + \text{H.c.}\right),
		\label{eq:h1}
	\end{equation}
	\begin{equation}
		H_{cd}= \delta \omega_{c} \sigma_{+}^{c}\sigma_{-}^{c} + \delta \omega_{d} \sigma_{+}^{d}\sigma_{-}^{d} + g_{{cd}} \left(\sigma_{+}^{d}\sigma_{-}^{c} + \text{H.c.}\right),
		\label{eq:h2}
	\end{equation}
are the Hamiltonian. Here, $\delta \omega_{j}$ and  $\sigma_{+}^j=|e_j\rangle \langle g_j|$($\sigma_{-}^j=|g_j\rangle \langle e_j|$) are the Lamb shift and raising (lowering) operator of atom $j$ $(j=a,b,c,d)$, $g_{{ab}}$ and $g_{{cd}}$ are exchange interaction for $ab$ and $cd$. In addition, in Eq. (\ref{eq:me1}), $\Gamma_{j}$ represents the individual decay rate of atom $j$, while $\Gamma_{ab}$ and $\Gamma_{cd}$ represent the collective decay rates for $ab$ and $cd$, respectively. The notation $D[O] \rho = O \rho O^\dagger - \frac{1}{2} \left( O^\dagger O \rho + \rho O^\dagger O \right)$ is used to describe the Lindblad superoperator governing atomic dissipation. For these coefficients in Eq. (\ref{eq:me1}), there is a unified expression applicable to the configurations shown in Figs. \ref{fig1}$(b)$–\ref{fig1}$(d)$ \cite{Kockum47}:

	\begin{equation}
		\begin{aligned}
			\delta \omega_{j_{{=a,b,c,d}}} &= \sum_{n=1}^{2} \sum_{m=1}^{2} \frac{\gamma}{2} \left[ \sin (\phi_{j_n,j_m}) \right] \\
			g_{{ab}} &= \sum_{n=1}^{2} \sum_{m=1}^{2} \frac{\gamma}{2} \left[ \sin (\phi_{a_n,b_m} ) \right] \\
			g_{{cd}} &= \sum_{n=1}^{2} \sum_{m=1}^{2} \frac{\gamma}{2} \left[ \sin (\phi_{c_n,d_m} ) \right] \\
			\Gamma_{j_{=a,b,c,d}} &= \sum_{n=1}^{2} \sum_{m=1}^{2} \gamma \left[\cos (\phi_{j_n,j_m}) \right] \\
			\Gamma_{ab} &= \sum_{n=1}^{2} \sum_{m=1}^{2} \gamma \left[\cos (\phi_{a_n,b_m}) \right] \\
			\Gamma_{cd} &= \sum_{n=1}^{2} \sum_{m=1}^{2} \gamma \left[\cos (\phi_{c_n,d_m}) \right] ,
		\end{aligned}
		\label{eq:ex3}
	\end{equation}
where $\gamma$ is the radiative decay rate of an atom at each connection point. It can be seen that these coefficients are functions of the phase shift. In Eq. (\ref{eq:ex3}), the term $\phi_{j_n,j_m}$ denotes the phase shift between the connection points $n$ and $m$ of atom $j$, and the other phase shift terms can be explained in a similar way.
For the phase shift-dependent coefficients expressions of the small atom configuration shown in Fig. \ref{fig1}$(a)$, it can be found in Refs. \cite{Lehmberg2, Kevin88, Pichler91} or derived by setting $n,m = 1$ in Eq. (\ref{eq:ex3}).

Note that Eq. (\ref{eq:me1}) is derived under the following assumptions: $(i)$ the coupling between the atoms and the waveguide is weak, and $(ii)$ the time taken for photons to travel through the distance $d_0$ between the connection points, which is much smaller than the atomic lifetime, is neglected.

\subsection{Non-Hermitian effective Hamiltonian and System dynamics evolution}
When only one excitation is allowed in the system, we follow the standard method described in Refs. \cite{Yin55, liu2024, Mok101} to derive the non-Hermitian effective Hamiltonian:
\begin{equation}
	\begin{aligned}
		\hat{H}_{\text{eff}} &= H_{ab} + H_{cd} - \frac{i}{2} \sum_{j=a,b,c,d} \Gamma_{j} \sigma_+^j \sigma_-^j \\
		&- \frac{i}{2} \Gamma_{ab} (\sigma_+^b \sigma_-^a + \sigma_+^a \sigma_-^b) \\
		&- \frac{i}{2} \Gamma_{cd} (\sigma_+^d \sigma_-^c + \sigma_+^c \sigma_-^d).
	\end{aligned}
	\label{Heff}
\end{equation}
For the four-qubit system considered in this work, the general state in single-excitation subspace can be expressed as
\begin{equation}
	\begin{aligned}
	\label{Fstate}
	|\psi(t)\rangle &= x_{1}(t) |e_a g_b g_c g_d \rangle +  x_{2}(t) |g_a g_b e_c g_d \rangle \\
	&+ x_{3}(t) |g_a e_b g_c g_d \rangle + x_{4}(t) |g_a g_b g_c e_d \rangle.
\end{aligned}
\end{equation}
Substituting Eqs. (\ref{Heff}) and (\ref{Fstate}) into the Schrödinger equation
\begin{equation}
	i \hbar \frac{\partial}{\partial t} |\psi(t)\rangle = \hat{H}_{\text{eff}} |\psi(t)\rangle,
\end{equation}
and according to the initial condition given in Eq. (\ref{Istate}), the analytical expressions for the probability amplitudes $x_{1}(t)$, $x_{2}(t)$, $x_{3}(t)$, and $x_{4}(t)$, which depend on the Lamb shift, exchange interaction strength, individual decay rates, and collective decay rate, can be derived.

\subsection{Entanglement measurement}

Since this work focuses only on the entanglement between pairs of atoms, the concurrence is adopted as the entanglement measure \cite{Wootters}. The concurrence is defined as follows:
\begin{equation}
C(\rho) = \max \left\{ 0, \sqrt{\lambda_1} - \sqrt{\lambda_2} - \sqrt{\lambda_3} - \sqrt{\lambda_4} \right\},
\end{equation}
where $\lambda_i$ ($i=1, 2, 3, 4$) are the eigenvalues of the non-Hermitian matrix $\rho (\sigma_y \otimes \sigma_y) \rho^* (\sigma_y \otimes \sigma_y)$ in descending order. Here, $\rho$ is the density matrix, $\rho^*$ is the complex conjugate of $\rho$, and $\sigma_y$ is the Pauli matrix.

In any two two-level atoms system, the density matrix is often expressed in the following form:
\begin{equation}
	\label{pmatrix}
	\rho(t) = \begin{pmatrix}
		H & 0 & 0 & W \\
		0 & I & Z & 0 \\
		0 & Z^*  & Q & 0 \\
		W^* & 0 & 0 & P
	\end{pmatrix},
\end{equation}
where $H + I + Q + P = 1$, $W^*$ and $Z^*$ are the complex conjugates of $W$ and $Z$, respectively.

Then, the concurrence can be simplified as \cite{Wootters}:
\begin{equation}
C = 2 \max \{ 0, |Z| - \sqrt{HP}, |W| - \sqrt{IQ} \},
\label{eq:ex2}
\end{equation}
the concurrence satisfies $0 \leq C \leq 1$, where $C = 0$ and $C = 1$ represent no entanglement and maximum entanglement, respectively.

\section{Entanglement transfer from atom pair \texorpdfstring{$ac$}{ac} to atom pair \texorpdfstring{$bd$}{bd}}\label{III}

This section focuses on the entanglement transfer from atom pair $ac$ to atom pair $bd$ in the four configurations illustrated in Fig. \ref{fig1}. For this purpose, we derive a general formula for the concurrence of atom pairs $ac$ and $bd$, applicable to the four configurations.

Tracing out the state associated with atoms $b$ and $d$ in Eq. (\ref{Fstate}), in the basis $\left\{|e\rangle_a |e\rangle_c, |e\rangle_a |g\rangle_c, |g\rangle_a |e\rangle_c, |g\rangle_a |g\rangle_c \right\}$, the reduced density matrix of atoms $a$ and $c$ can be determined

\begin{align}
	\label{pmatrix1}
	\rho^{ac}(t) &= \mathrm{Tr}_{bd} \rho(t) = \mathrm{Tr}_{bd} (|\psi(t)\rangle \langle \psi(t)|) \notag \\
	&= \begin{pmatrix}
		0 & 0 & 0 & 0 \\
		0 & |{x}_{1}(t)|^2 & {x}_{1}(t) {x}_{2}^*(t) & 0 \\
		0 & {x}_{1}^*(t) {x}_{2}(t) & |{x}_{2}(t)|^2 & 0 \\
		0 & 0 & 0 & |{x}_{3}(t)|^2+|{x}_{4}(t)|^2
	\end{pmatrix}.
\end{align}

According to Eq. (\ref{eq:ex2}), the concurrence of atom pair $ac$ is described as:
\begin{equation}
	C_{ac}(t) =2|x_{1}(t) x_{2}^{\ast }(t)|.\label{Concurrence1}
\end{equation}

Likewise, the reduced density matrix of the system formed by atoms $b$ and $d$ in the basis $\left\{|e\rangle_b |e\rangle_d, |e\rangle_b |g\rangle_d, |g\rangle_b |e\rangle_d, |g\rangle_b |g\rangle_d \right\}$ can be obtained as:
\begin{align}
	\label{pmatrix2}
	\rho^{bd}(t) &= \mathrm{Tr}_{ac} \rho(t) = \mathrm{Tr}_{ac} (|\psi(t)\rangle \langle \psi(t)|) \notag \\
	&= \begin{pmatrix}
		0 & 0 & 0 & 0 \\
		0 & |{x}_{3}(t)|^2 & {x}_{3}(t) {x}_{4}^*(t) & 0 \\
		0 & {x}_{3}^*(t) {x}_{4}(t) & |{x}_{4}(t)|^2 & 0 \\
		0 & 0 & 0 & |{x}_{1}(t)|^2+|{x}_{2}(t)|^2
	\end{pmatrix}.
\end{align}
Consequently, the entanglement between atoms $b$ and $d$ can be measure as
\begin{equation}
	C_{bd}(t) =2|x_{3}(t) x_{4}^{\ast }(t)|.\label{Concurrence2}
\end{equation}

In the following text, the superscripts $S$, $B$, and $N$ in the concurrence refer to the separated, braided, and nested configurations, respectively, while the superscript $s$ represents the small atom configuration.

\subsection{Entanglement transfer in small atom configuration}\label{III A}
We first investigate the entanglement transfer in small atom configuration. We note that Ref. \cite{Cavalcanti74} reported entanglement transfer between two pairs of small atoms without considering dissipation. Here, we address the case where dissipation is taken into account during entanglement transfer. In this configuration, as illustrated in Fig. \ref{fig1}($a$), the Lamb shift, exchange interaction strength, individual decay rates, and collective decay rate are given by the following expressions:
\begin{equation}
	\begin{aligned}
		\delta \omega_{j_{{=a,b,c,d}}}&= 0 \\
		g_{ab} =g_{cd} &= \frac{\gamma}{2}  \sin \phi  \\
		\Gamma_{j_{{=a,b,c,d}}}&= \gamma \\
		\Gamma_{ab}=\Gamma_{cd}&=\gamma   \cos \phi .
	\end{aligned}
	\label{eq:ex21}
\end{equation}
\begin{figure}[!htbp]
	\includegraphics[width=0.40\textwidth]{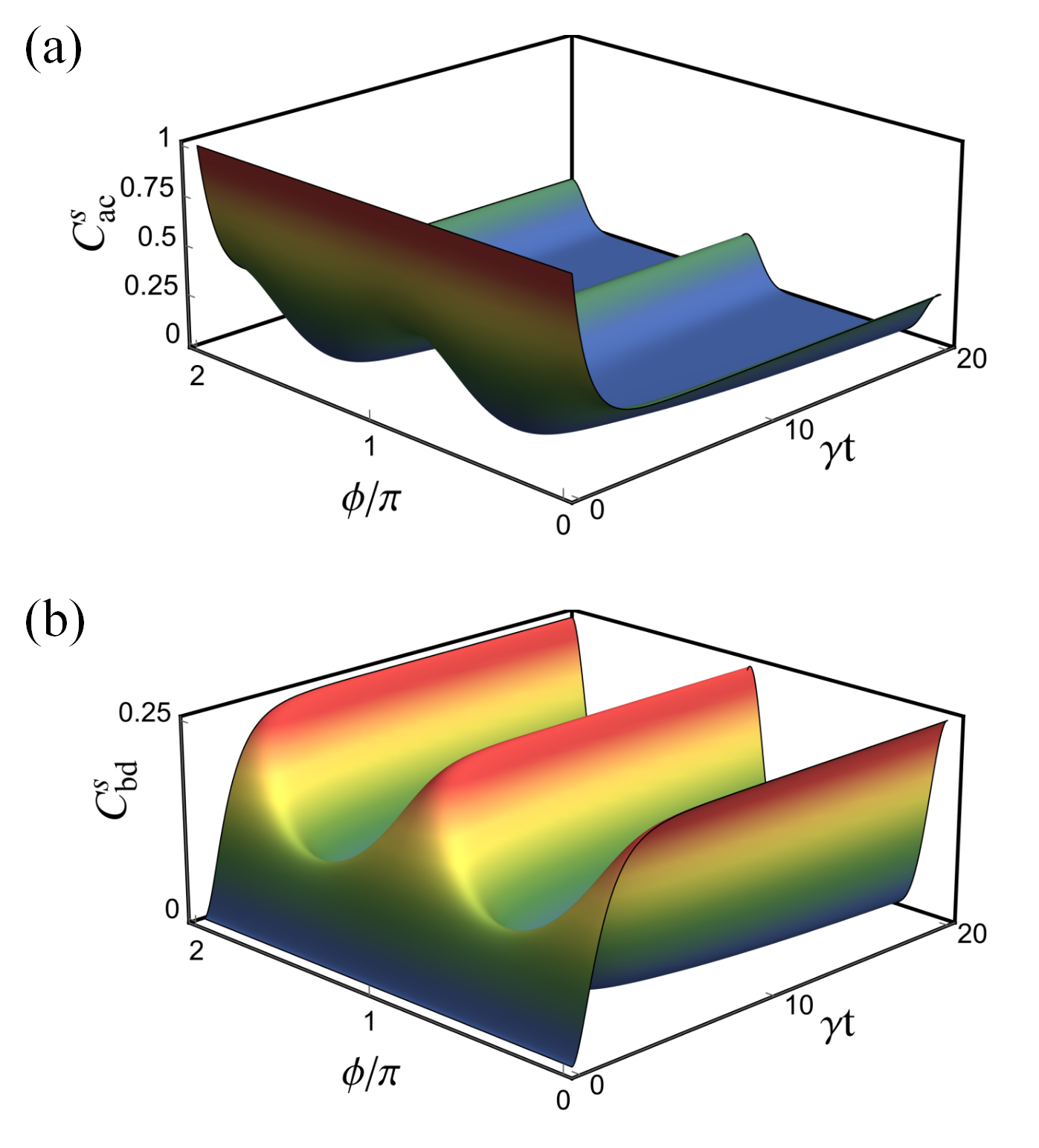}\hfill
	\caption{(Color online)  Concurrences $(a)$ $C_{ac}^{s}$ and $(b)$ $C_{bd}^{s}$ as a function of $\phi/\pi$ and $\gamma t$. }
	\label{fig2}
\end{figure}
In Figs. \ref{fig2}($a$) and \ref{fig2}($b$), the concurrences $C_{ac}^{s}$ and $C_{bd}^{s}$, respectively, are shown as a function of $\phi/\pi$ and $\gamma t$. It is evident from Figs. \ref{fig2}($a$) and \ref{fig2}($b$) that the concurrences $C_{ac}^{s}$ and $C_{bd}^{s}$ are both influenced by the phase shift $\phi $. Additionally, the dependence of the two concurrences on $\phi $ demonstrates a periodic behavior with a period $\pi$. 

From Fig. \ref{fig2}($a$), we can see that the concurrence $C_{ac}^{s}$ of the atom pair $ac$, which are initially in the maximally value, then rapidly decays to 0 within a short time for $\phi \in [0, 2\pi]$, except when $\phi$ = $0, \pi, 2\pi$, where it remains at a steady state value 0.25. For the purpose of explaining the steady-state behavior of the concurrence $C_{ac}^{s}$, the analytical formulas for the concurrences under these phase shifts are calculated. Here, we take $\phi$ = $0$ as an example to illustrate the concurrence expression. When $\phi$ = $0$, the concurrence is given by

\begin{equation}
	C_{ac}^{s}(t)=\frac{1}{4}  \left(1 + 2e^{- \gamma t} + e^{-2\gamma t}\right).
	\label{eq:C1}
\end{equation}
In Eq. (\ref{eq:C1}), we find that the concurrence $C_{ac}^{s}(t)$ asymptotically tends to a steady-state value of 0.25 as time goes. In addition, in comparison to $2e^{-\gamma t}$, the influence of the term $e^{-2\gamma t}$ on the concurrence can be neglected because it decays significantly faster with increasing $\gamma t$. Consequently, at $\phi = 0$, the rate at which the concurrence $C_{ac}^{s}(t)$ reaches the steady-state value 0.25 is $\gamma$. From here on, we will concentrate on the terms that contribute more significantly to the concurrence and ignore the rapidly decaying ones. 

For the concurrence $C_{bd}^{s}$, shown in Fig. \ref{fig2}($b$), we observe that it increase gradually with $\gamma t$ at $\phi$ = $0, \pi, 2\pi$ and eventually reaches a steady-state value 0.25. An explanation for this behavior is as follows: initially, the atom pair $bd$ is in the ground state, which corresponds to a separable state. As $\gamma t$ increases, the entanglement in atom pair $ac$ gradually transfers to atom pair $bd$ via the waveguide. However, due to the influence of dissipation, the entanglement transfer from atom pair $ac$ to atom pair $bd$ is finite. When $\gamma t$ further increases, the concurrence $C_{bd}^{s}$ no longer increases and remains at the steady state value 0.25. On the other hand, this can also be explained using the analytical expression of the concurrence $C_{bd}^{s}$.
We here also give the analytical form of $C_{bd}$ for $\phi$ = $0$ as an example: 
\begin{equation}
	C_{bd}^{s}(t)=\frac{1}{4}  \left(1 - 2e^{- \gamma t} + e^{-2\gamma t}\right).
	\label{eq:C6}
\end{equation}

From Eq. (\ref{eq:C6}), we can find that at $\gamma t = 0$, the concurrence $C_{bd}^{s}(t)$ = 0, and as $\gamma t$ increases, $C_{bd}^{s}(t)$ approaches a steady-state value 0.25 with a rate $\gamma$.

\subsection{Entanglement transfer in separated configuration}\label{III B}
We now discuss the entanglement transfer in separated configuration. Based on the arrangement of the giant atom and waveguide coupling points shown in Fig. \ref{fig1}($b$), as well as the formula given in Eq. (\ref{eq:ex3}), we derive the expressions for the Lamb shift, exchange interaction strength, individual decay rates, and collective decay rate as a function of the phase shift $\phi$, as follows:
\begin{equation}
	\begin{aligned}
		\delta \omega_{j_{{=a,b,c,d}}}&= \gamma \sin \phi  \\
		g_{ab}=g_{cd} &= \frac{\gamma}{2} \left[ \sin \phi + 2\sin 2 \phi + \sin 3\phi \right] \\
		\Gamma_{j_{{=a,b,c,d}}}&= 2\gamma + 2\gamma \cos \phi  \\
		\Gamma_{ab}=\Gamma_{cd}&=\gamma \left[  \cos\phi + 2\cos 2\phi + \cos 3\phi \right].
	\end{aligned}
	\label{eq:ex22}
\end{equation}

\begin{figure}[!htbp]
	\includegraphics[width=0.40\textwidth]{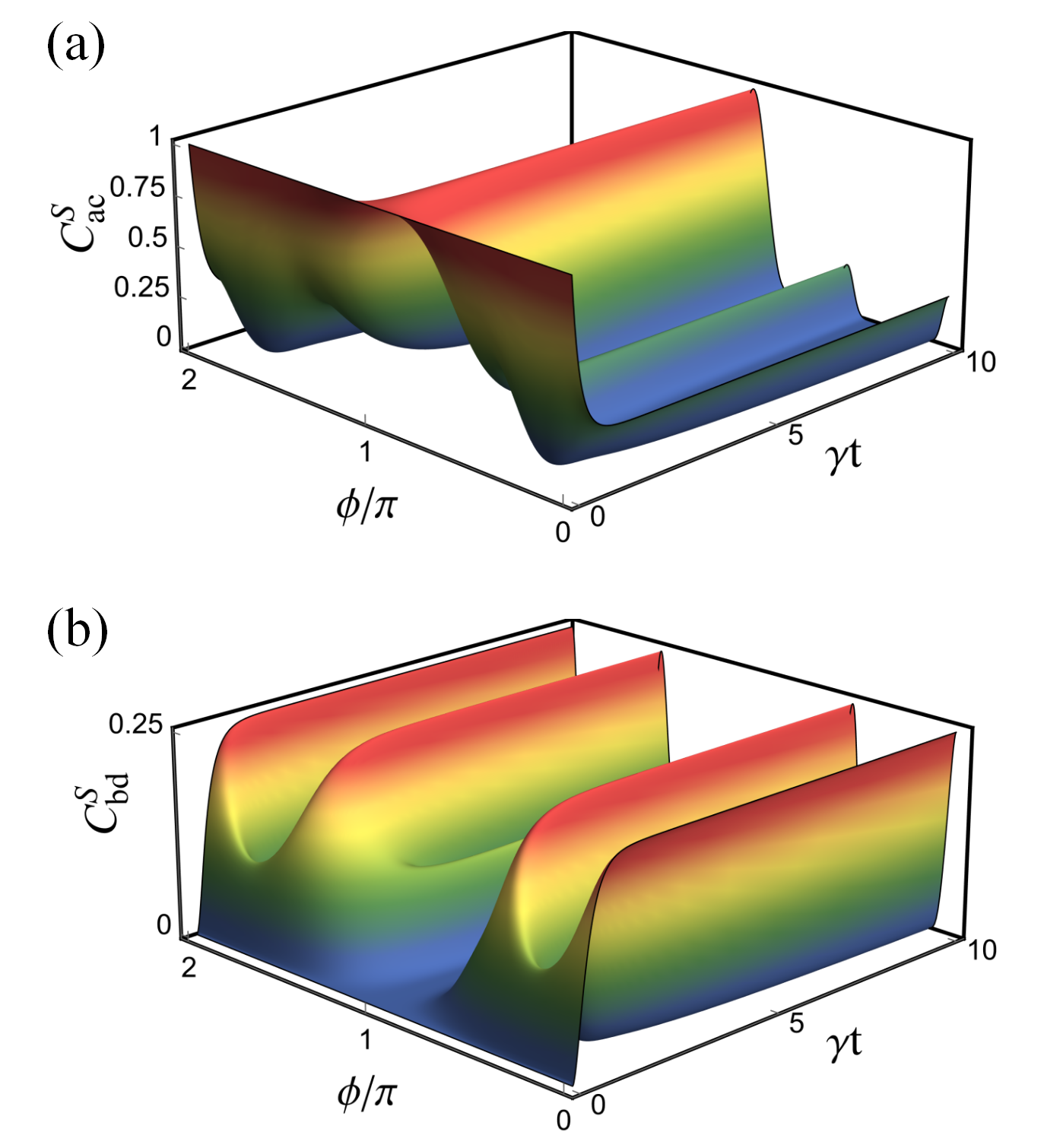}\hfill
	\caption{(Color online)  Concurrences $(a)$ $C_{ac}^{S}$ and $(b)$ $C_{bd}^{S}$ as a function of $\phi/\pi$ and $\gamma t$. }
	\label{fig3}
\end{figure}

For the separated configuration, the concurrences $C_{ac}^{S}$ and $C_{bd}^{S}$ of atom pairs $ac$ and $bd$, as a function of $\phi/\pi$ and $\gamma t$, are depicted in Fig. \ref{fig3}($a$) and Fig. \ref{fig3}($b$), respectively. Unlike the small atom configuration, the phase shift $\phi$ governs the concurrences $C_{ac}^{S}$ and $C_{bd}^{S}$ with a period $2\pi$.

In Fig. \ref{fig3}($a$), we can see that the concurrence $C_{ac}^{S}$ does not change as time goes, keeping $C_{ac}^{S}$ = 1 when $ \phi$ is taken as $ \pi$. This is mainly due to the fact that the coefficients in Eq. (\ref{eq:ex22}) are all zero when $ \phi$ = $ \pi$. This means that the atom pairs $ab$ and $cd$ become decoupled from the waveguides to which they are coupled, so that the initial maximally entangled quantum state of the atom pair $ac$ will always remain and does not transfer to the atom pair $bd$. Numerically, this is demonstrated as the concurrence $C_{ac}^{S}=1$ for all times. Correspondingly, as illustrated in Fig. \ref{fig3}($b$), there is no entanglement at $\phi$ = $\pi$, i.e., $C_{bd}^{S}=0$ with time evolution.
At other $ \phi $, such as $ \phi$ = $ 0, \pi/2, 3\pi/2, 2\pi$, the concurrence $C_{ac}^{S}$ (Fig. \ref{fig3}($a$)) undergo decay over time, whereas $C_{bd}^{S}$ (Fig. \ref{fig3}($b$)) increases with time. When $\gamma t$ is sufficiently large, both concurrences $C_{ac}^{S}$ and $C_{bd}^{S}$ eventually converge to a steady state value 0.25. This behavior is similar to that exhibited in the small atom configuration. However, the rates at which $C_{ac}^{S}$ and $C_{bd}^{S}$ reach the steady-state value differ from those in the small atom configuration. For example, when $ \phi$ =  0, the analytical expressions for the concurrences $C_{ac}^{S}$ and $C_{bd}^{S}$ are given as
\begin{equation}
	C_{ac}^{S}(t)=\frac{1}{4}  \left(1 + 2e^{- 4\gamma t} + e^{-8\gamma t}\right),
\end{equation}
\begin{equation}
	C_{bd}^{S}(t)=\frac{1}{4}  \left(1 - 2e^{- 4\gamma t} + e^{-8\gamma t}\right).
	\label{eq:C8}
\end{equation}

Hence, the convergence rate of $C_{ac}^{S}(t)$ and $C_{bd}^{S}(t)$ to the steady-state value 0.25 is $4\gamma$, which is four times the rate at which $C_{ac}^{s}(t)$ and $C_{bd}^{s}(t)$ approach the steady-state value. Moreover, a comparison between Fig. \ref{fig2} and Fig. \ref{fig3} reveals that the phase shift positions and their number, where steady-state entanglement is generated, differ between the small atom and separated configurations. The difference arises mainly from the distinct quantum interference effects between the coupling points in the two configurations.

\subsection{Entanglement transfer in braided configuration }\label{III C}
We now turn our attention to the entanglement transfer in braided configuration. In this case, the Lamb shift, exchange interaction strength, individual decay rates, and collective decay rate can be obtained as
\begin{equation}
	\begin{aligned}
		\delta \omega_{j_{{=a,b,c,d}}}&= \gamma \sin 2\phi  \\
		g_{ab}=g_{cd} &= \frac{\gamma}{2} \left[ 3\sin \phi + \sin 3\phi \right] \\
		\Gamma_{j_{{=a,b,c,d}}}&= 2\gamma + 2\gamma \cos 2\phi  \\
		\Gamma_{ab}=\Gamma_{cd}&=\gamma \left[  3\cos\phi +  \cos 3\phi \right].
	\end{aligned}
	\label{eq:ex26}
\end{equation}
From Eq. (\ref{eq:ex26}), we find that for phase shifts $\phi$= $ \pi/2$ or $ 3\pi/2$, the giant atoms $a$ ($c$) and $b$ ($d$) exhibit interaction without decoherence \cite{Kockum47}. Such a feature is not found in the other configurations addressed in this paper. Moreover, as will be shown below, this interaction without decoherence results in significant differences in the entanglement transfer behavior between the braided configuration and other configurations.

\begin{figure}[!htbp]
	\includegraphics[width=0.40\textwidth]{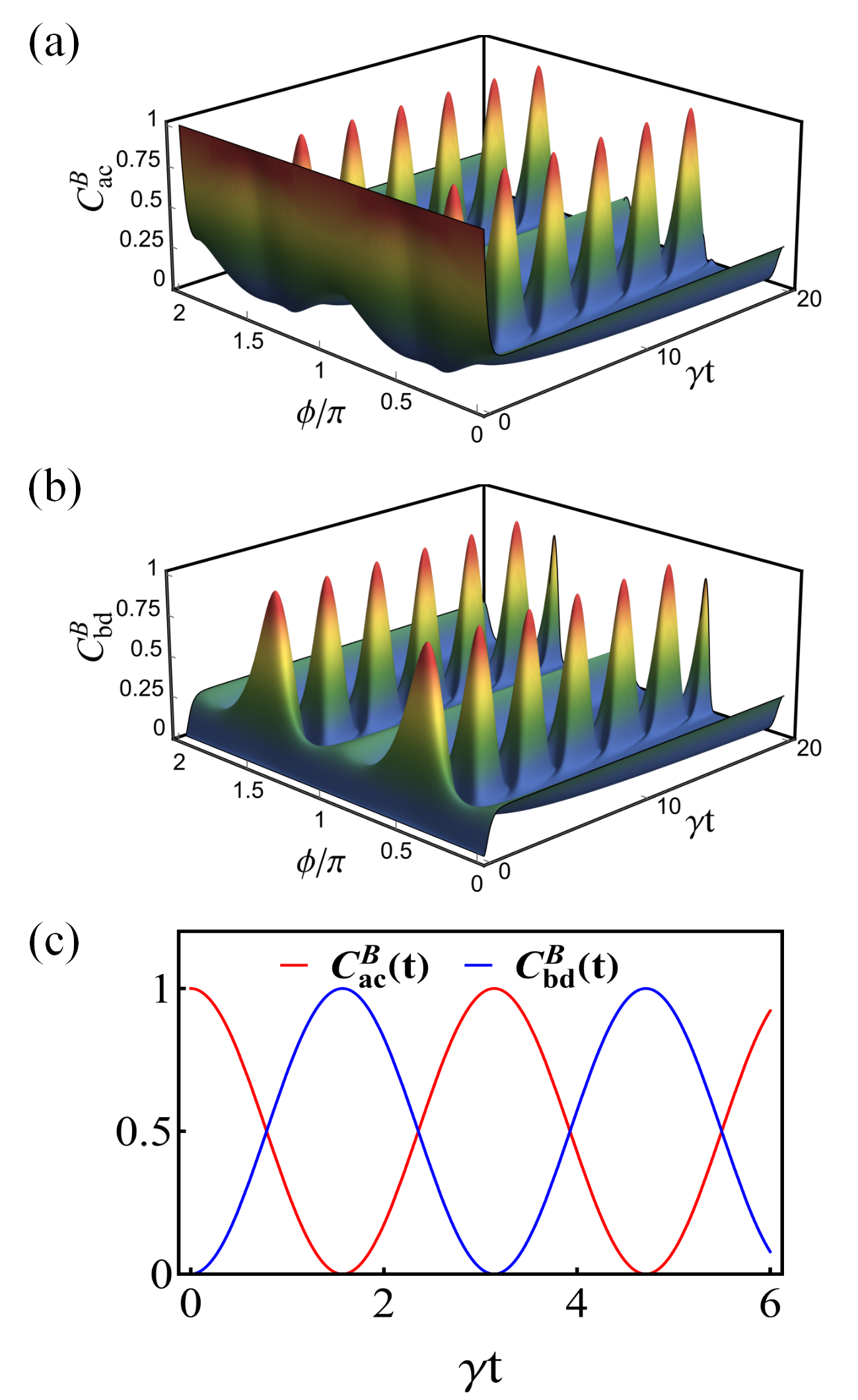}\hfill
	\caption{(Color online)  Concurrences $(a)$ $C_{ac}^{B}$ and $(b)$ $C_{bd}^{B}$ as a function of $\phi/\pi$ and $\gamma t$, $(c)$ $C_{ac}^{B}(t)$ and $C_{bd}^{B}(t)$ as a function of $\gamma t$ for a given value of $\phi$.}
	\label{fig4}
\end{figure}

In Figs. \ref{fig4}($a$) and \ref{fig4}($b$), we show the evolution of concurrences $C_{ac}^{B}$ and $C_{bd}^{B}$ as a function of $\phi/\pi$ and $\gamma t$. From Figs. \ref{fig4}($a$) and \ref{fig4}($b$), we see that the control of the concurrences $C_{ac}^{B}$ and $C_{bd}^{B}$ by the phase shift $\phi$ is periodic with a period of $\pi$.

Meanwhile, in Figs. \ref{fig4}($a$) and \ref{fig4}($b$), it can be observed that the concurrences $C_{ac}^{B}$ and $C_{bd}^{B}$ exhibit steady-state characteristics at $ \phi$ = $ 0, \pi, 2\pi$, which is analogous to the small atom configuration case. For phase shift $ \phi$ = $ \pi/2$ and 3$\pi/2$, both concurrences $C_{ac}^{B}$ (Fig. \ref{fig4}($a$)) and $C_{bd}^{B}$ (Fig. \ref{fig4}($b$)) periodic oscillate between 0 and 1, highlighting that the entanglement in atom pair $ac$ can be completely transferred to atom pair $bd$. To better observe the process of entanglement transfer from atom pair $ac$ to atom pair $bd$, we plot the concurrence as a function of $\gamma t$ in Fig. \ref{fig4}($c$), where $ \phi$ = $ \pi/2$. 
It can be seen from Fig. \ref{fig4}($c$), initially, the atom pair $ac$ exhibit maximum entanglement, while no entanglement is present between giant atoms $b$ and $d$. As the concurrence $C_{ac}^{B}(t)$ decreases, there is a corresponding increase in the concurrence $C_{bd}^{B}(t)$. When the entanglement of atom pair $ac$ becomes 0, the concurrence $C_{bd}^{B}(t)$ of atom pair $bd$ reaches its maximum value 1, realizing maximum entanglement transfer. This behavior can be attributed to the following physical mechanism: when $ \phi$ = $ \pi/2$, the dissipation terms in Eq. (\ref{eq:ex26}) vanishes. Nevertheless, the exchange interaction strengths $g_{ab}$ and $g_{cd}$ of the atom pairs $ab$ and $cd$, mediated by the waveguide, remain nonzero, which allows the decoherence-free transfer of the entanglement from the atom pair $ac$ to atom pair $bd$.
We also can obtained the analytical expressions for the concurrences $C_{ac}^{B}$ and $C_{bd}^{B}$ at the phase shift $ \phi$ = $ \pi/2$. In this phase shift value, the concurrences are calculated as 
\begin{equation}
C_{ac}^{B}(t)=\cos^2(\gamma t) ,
\label{acB}
\end{equation}
\begin{equation}
	C_{bd}^{B}(t)=\sin^2(\gamma t). 
	\label{bdB}
\end{equation}
As a result, $C_{ac}^B(t)$ and $C_{bd}^B(t)$ exhibit periodic oscillations between 0 and 1. Moreover, as can be seen from Eq. (\ref{acB}) and Eq. (\ref{bdB}), the phase shift difference between $C_{ac}^B(t)$ and $C_{bd}^B(t)$ is $\pi$, which means that when $C_{ac}^B(t)$ reaches its maximum value, $C_{bd}^B(t)$ reaches its minimum value, and vice versa. This two equations reflects the periodic transfer behaviour of entanglement between giant atoms pairs $ac$ and $bd$. When the phase shift $\phi$ = $3\pi/2$, a similar explanation holds, and thus, further discussion is omitted here.

\subsection{Entanglement transfer in nested configuration}\label{III D}
Finally, we focus on the entanglement transfer between giant atoms in nested configuration. For the configuration shown in Fig. \ref{fig1}($d$), the expressions for the coefficients, which depend on the phase shift $\phi$, are as follows:
\begin{equation}
	\begin{aligned}
		\delta \omega_{a}=\delta \omega_{c}&= \gamma \sin 3\phi  \\
		\delta \omega_{b}=\delta \omega_{d}&= \gamma \sin \phi  \\
		g_{ab}=g_{cd} &= \gamma \left[ \sin \phi + \sin 2\phi \right] \\
		\Gamma_{a}=\Gamma_{c}&= 2\gamma + 2\gamma \cos 3\phi  \\
		\Gamma_{b}=\Gamma_{d}&= 2\gamma + 2\gamma \cos \phi  \\
		\Gamma_{ab}=\Gamma_{cd}&=2\gamma \left[  \cos\phi +  \cos 2\phi \right].
	\end{aligned}
	\label{eq:ex24}
\end{equation}

\begin{figure}[!htbp]
	\includegraphics[width=0.40\textwidth]{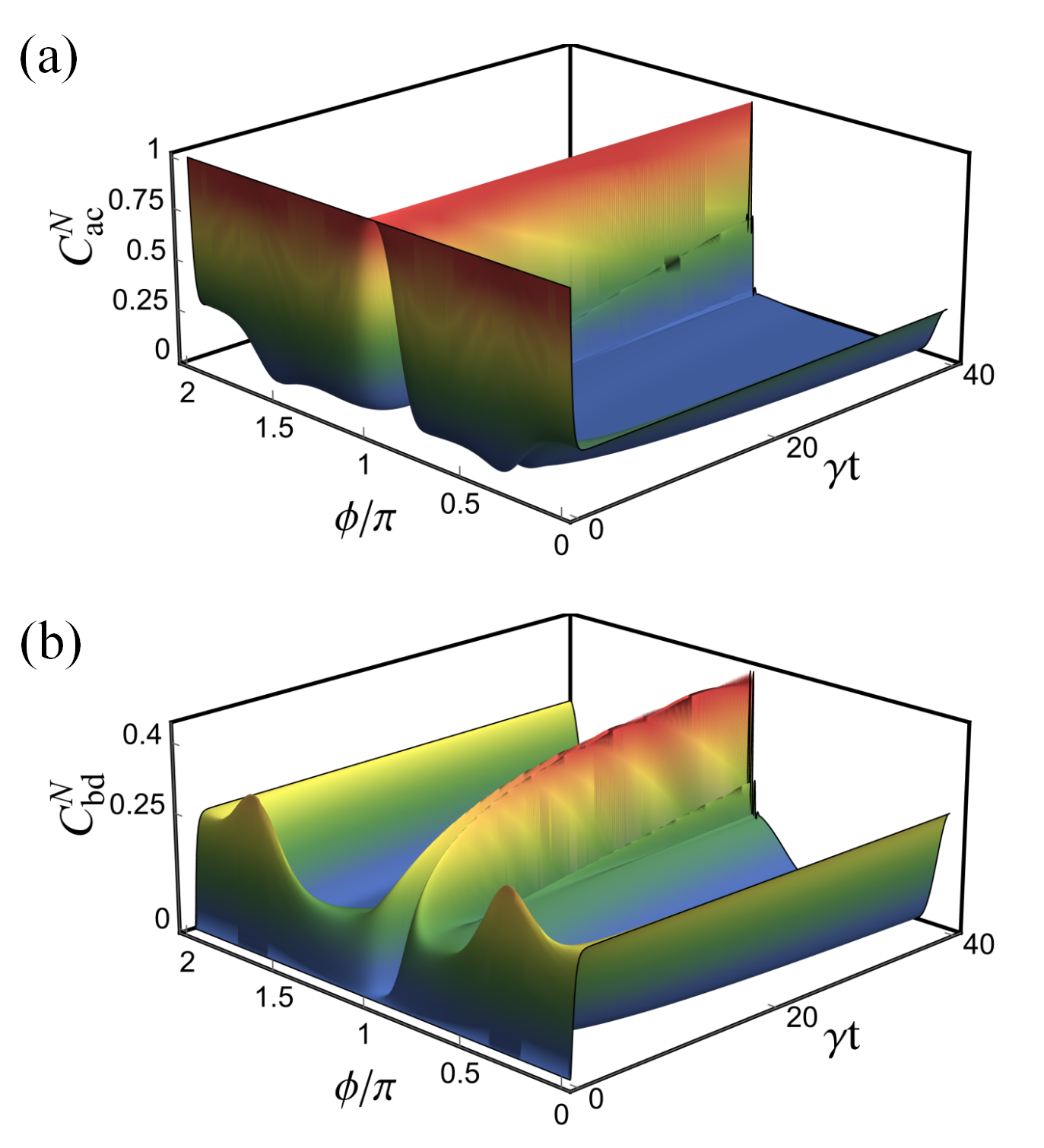}\hfill
	\caption{(Color online)  Concurrences $(a)$ $C_{ac}^{N}$ and $(b)$ $C_{bd}^{N}$ as a function of $\phi/\pi$ and $\gamma t$.}
	\label{fig5}
\end{figure}

In Figs. \ref{fig5}($a$) and \ref{fig5}($b$), we show the concurrences $C_{ac}^N$ and $C_{bd}^N$, respectively, as a function of $\phi/\pi$ and $\gamma t$. Consistent with the behavior observed in the separated configuration, concurrences $C_{ac}^N$ and $C_{bd}^N$ demonstrate a phase shift dependence with a period $2\pi$. 
When $\phi = \pi$, the concurrence $C_{ac}^N$ remains unchanged at its maximum value 1 over time, and correspondingly, $C_{bd}^N = 0$, as shown in Figs. \ref{fig5}($a$) and \ref{fig5}($b$). This behavior shares a common physical mechanism with that exhibited in separated configuration. Furthermore, concurrences $C_{ac}^N$ and $C_{bd}^N$ approaches a steady state value 0.25 at $\phi = 0, 2\pi$. Such characteristics are similar to those demonstrated in the other configurations. In this configuration, however, the peak value of entanglement transfer is greater than that in the small atom and separated configurations. As shown in Fig. \ref{fig5}($b$), concurrence $C_{bd}^N$ exhibits two ridges at $\phi \rightarrow \pi$, and the value of concurrence $C_{bd}^N$ increase with the extension of $\gamma t$. When $\phi \rightarrow \pi$, the peak value of $C_{bd}^N \approx 0.42$, a value that is greater than the entanglement transfer peak value 0.25 in the small atom and separated configurations. In addition, a sub-peak with a value 0.33 is found at the phase shift $\phi = \pi/3$ and $\phi = 5\pi/3$.

\section{Entanglement transfer from atom pair \texorpdfstring{$ac$}{ac} to atom pair \texorpdfstring{$ab$}{ab}}\label{IV}
In Sec. \ref{III}, we studied the transfer of entanglement from atom pair $ac$ to atom pair $bd$. The entanglement in atom pair $ac$ can be transferred to other atom pairs, such as atom pair $ab$, in addition to atom pair $bd$. In this section, we analyze the entanglement transfer from atom pair $ac$ to atom pair $ab$ in the four configurations previously described. Since the entanglement evolution of atom pair $ac$ is given in Sec. \ref{III}, here we only present the entanglement dynamic for atom pair $ab$. Repeating the procedure described in ~Sec. \ref{III}, the concurrence of atom pair $ab$ takes the following form:
\begin{equation}
	C_{ab}(t) =2|x_{1}(t) x_{3}^{\ast }(t)|.\label{Concurrence3}
\end{equation}
\begin{figure*}[!htbp]
	\includegraphics[width=0.98\textwidth]{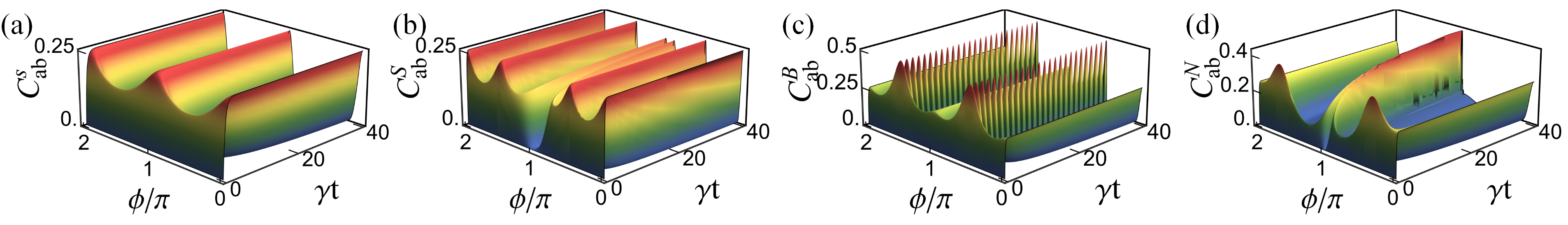}\hfill
	\caption{(Color online)  Concurrences $(a)$ $C_{ab}^{s}$, $(b)$ $C_{ab}^{S}$, $(c)$ $C_{ab}^{B}$ and $(d)$ $C_{ab}^{N}$ as a function of $\phi/\pi$ and $\gamma t$.}
	\label{fig6}
\end{figure*}
In Figs. \ref{fig6}($a$)- \ref{fig6}($d$), we plot the evolution of concurrence with phase shift $\phi/\pi$ and $\gamma t$ for small atom, separated, braided, and nested configurations. For small atom and separated configurations, as illustrated in Figs. \ref{fig6}($a$) and \ref{fig6}($b$), concurrences $C_{ab}^s$ and $C_{ab}^S$ have similar steady-state entanglement properties to concurrences $C_{bd}^s$ and $C_{bd}^S$, as can be seen by comparing with Figs. \ref{fig2}($b$) and \ref{fig3}($b$). It should be noted that while $C_{ab}^s$ and $C_{ab}^S$ can also reach the steady-state value of 0.25, the rate at which they do so is different from that of $C_{bd}^s$ and $C_{bd}^S$. For example, when $\phi = 0$, we have
\begin{equation}
	C_{ab}^s(t)=\frac{1}{4} \left( 1 - e^{-2\gamma t} \right),\label{Concurrenceab}
\end{equation}
\begin{equation}
C_{ab}^S(t)=\frac{1}{4} \left( 1 - e^{-8\gamma t} \right).\label{ConcurrenceSab}
\end{equation}
Therefore, the concurrences $C_{ab}^s(t)$ and $C_{ab}^S(t)$ approach the steady-state values at rates $2\gamma$ and $8\gamma$, respectively. Comparing Eqs. (\ref{Concurrenceab})- (\ref{ConcurrenceSab}) with Eq. (\ref{eq:C6}) and Eq. (\ref{eq:C8}), it can be seen that concurrences $C_{ab}^s(t)$ and $C_{ab}^S(t)$ attain the steady-state value at a faster rate. In the separated configuration, particularly, the concurrence $C_{ab}^S$ = 0 when $\phi = \pi$. This behavior is governed by the same physical mechanism as that for $C_{bd}^S$ = 0 when $\phi = \pi$.

For the braided configuration, concurrence $C_{ab}^B(t)$ at $\phi = 0, \pi, 2\pi$ asymptotically converges to the steady-state value 0.25, as shown in Fig. \ref{fig6}($c$). This behavior agrees with what is seen in the concurrence $C_{bd}^B$. However, for $\phi = \pi/2$ and $3\pi/2$, unlike the concurrence $C_{bd}^B$, which exhibits oscillatory behavior ranging from 0 to 1, concurrence $C_{ab}^B$ oscillates periodically only within the range from 0 to 0.5. The reason behind this behavior can be found in the analytical expression of concurrence $C_{ab}^B$ below. According to Eq. (\ref{Concurrence3}), when $\phi = \pi/2$ or $3\pi/2$, we have
\begin{equation}
C_{ab}^B(t)=\frac{1}{2} |\sin(2\gamma t)|.\label{ConcurrenceBab}
\end{equation}
From Eq. (\ref{ConcurrenceBab}), we find that $C_{ab}^B(t)$ shows periodic oscillations between 0 and 0.5, characterized by an oscillation period of \(\frac{\pi}{2\gamma}\). Consequently, in the braided configuration, when we consider transferring the entanglement in atom pair $ac$ to atom pair $ab$, the maximum entanglement value attainable for atom pair $ab$ is 0.5.

In Fig. \ref{fig6}($d$), the entanglement evolution of atom pair $ab$ is shown for the nested configuration. It can be observed that the concurrences $C_{ab}^N$ and $C_{bd}^N$ (Fig. \ref{fig5}($b$)) exhibit the similar evolution characteristics when $\phi = 0, \pi, 2\pi$ and $\phi \rightarrow \pi$. However, when $\phi = \pi/3$, the concurrence $C_{ab}^N \approx 0.39$, which is higher than that of concurrence $C_{bd}^N$.

\section{DISCUSSION}\label{V}
In this work, we specifically discuss the entanglement transfer from atom pair $ac$ to atom pair $bd$ and atom pair $ab$ in small atom and three giant atoms configurations. The transfer of entanglement from the atom pair $ac$ to other atom pairs, such as $cd$, $ad$, and $bc$, is not discussed here. This is due to our numerical result showing that, in each configuration, the concurrences $C_{cd} = C_{ad} = C_{bc} = C_{ab}$. 

Furthermore, in the previous sections, we only considered the case where the atom pair $ac$ is initially in the entangled state $({\left| e_a, g_c \right\rangle + \left| g_a, e_c \right\rangle})/{\sqrt{2}}$. 
For another single-excitation maximally entangled state $({\left| e_a, g_c \right\rangle - \left| g_a, e_c \right\rangle})/{\sqrt{2}}$, the behavior of entanglement transfer across the four coupling configurations is similar to the results presented in this paper; thus, the details are not included here for brevity. Note that in the braided configuration, the entanglement between giant atoms $a$ and $c$ can still be completely transferred to atom pair $bd$ in the initial condition $|\psi(0)\rangle = \frac{1}{\sqrt{2}} (| e_a, g_c \rangle - | g_a, e_c \rangle) \otimes | g_b, g_d \rangle$. This is attributed to the decoherence-free nature of the braided giant atom, which is robust against the atomic quantum states \cite{Kockum47}. 

Following, we analyze the experimental feasibility of the scheme proposed. Various platforms have successfully realized giant atoms, including transmon qubits coupled to surface acoustic waves \cite{Gustafsson2014, Manenti2017} and microwave transmission lines \cite{Kannan48, Vadiraj2021}, and ferromagnetic spin ensembles coupled with microwave transmission lines \cite{Wang2022}. Thus, our proposal is achievable within the current experimental technological.
\section{CONCLUSION}\label{VI}
In summary, this work investigates the entanglement transfer between giant atoms in waveguide-QED systems. Concretely, we mainly focus on the transfer of entanglement from atom pair $ac$ to atom pair $bd$ and atom pair $ab$ in three giant atom configurations, where the atom pair $ac$ initially prepared in a maximally entangled state. We also analyzed the entanglement transfer between two pairs of small atoms (i.e., small atom configuration). The research shows that the entanglement transfer between atom pairs depends on the phase shift and coupling configuration. 
Moreover, the maximum entanglement value transferred from atom pair $ac$ to atom pair $bd$ and atom pair $ab$ in the braided configurations is higher than that in the small atom configuration, separated configuration, and nested configuration. In particular, in the braided configuration, the entanglement stored in atom pair $ac$ can be completely transferred to atom pair $bd$. This work may contribute to advancing quantum information processing built on giant atom waveguide-QED systems.

\section*{ACKNOWLEDGMENTS}
This work was supported by National Natural Science Foundation of China(Grants No. 11874190 and No. 12247101). Support was also provided by Supercomputing Center of Lanzhou University.

\bibliography{REF}

\end{document}